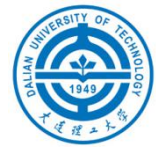

# Academic Lobification: Low-performance Control Strategy for Long-planed Academic Purpose

Shudong YANG[1]

(1. Dalian University of Technology, Dalian, China, 116024)

**Abstract**: Academic lobification refers to a collection of academic performance control strategies, methods, and means that a student deliberately hides academic behaviors, or deliberately lowers academic performance, or deliberately delays academic returns for a certain long-term purpose, but does not produce academic risks. Understanding academic lobification is essential to our ability to compensate for inherent deviations in the evaluation of students' academic performance, discover gifted student, reap benefits and minimize harms. It outlines a set of questions that are fundamental to this emerging interdisciplinary research field, including research object, research question, research scope, research method, and explores the technical, legal and other constraints on the study of academic lobification.

# 1. Introduction

Human beings are truly socialized animals. Lies are a common phenomenon in human society. Some lies are for expressing goodwill, some are for maintaining order, and some are for obtaining more long-term benefits. Lies more or less play a role in the construction of social network subject relationships[1]. When the social network expands to the two sides of the hostile, it also presents an infinitely attractive strategy for defeating the enemy and the wisdom of war[2].

Strategies exist in many social fields. For instance, **in the field of sports events**, ancient China had Tianji horse racing. In contemporary world-class competitions, generally speaking, the players are evenly matched. When they lack absolute strength advantages, they often adopt strategic strategies to win, such as: players hide their strength in the preliminaries in order to confuse their opponents[3]. When the player has no remaining claim rights, but has complete remaining control rights, the player will slow down the progress of each time in order to maximize the benefits. for example, Sergey Bubka have broken men's pole vault world record for 35 times. **In the political field**, politicians use fox-like methods to achieve the so-called lion-like goals hidden in his heart that have never been known to the outside world[4]. **In the business field**, hungry marketing strategies are suitable for high-tech startups to release flagship products, such as Xiaomi mobile phones[5]. Some high-tech companies squeeze technology like toothpaste at will[6]. For individuals, in a scenario where performance standards increase with performance improvement, rational people deliberately hide their strength in order to avoid the

[1] Jia Jia. The Truth and Falsehooel of "Lies" ——A Semiotic Proposition[J]. Journal of Hebei Normal University (Philosophy and Social Sciences Edition), 2020,43(5):13-19. DOI:10.13763/j.cnki.jhebnu.psse.2020.05.003.
[2] Yan Shengguo. On Strategy and Wisdom to Conquer the Enemy in the Art of War[J]. Journal of Henan Normal University (Philosophy and Social Sciences), 2020,47(2):82-88. DOI:10.16366/j.cnki.1000-2359.2020.02.012.
[3] Li Yiqun, Xie Yalong. Sports Game Theory[M]. Beijing Sport University Press, 2002: 63, 210, 214
[4] James Mac Gregor Burns. Roosevelt : the lion and the fox[M]. Harcourt Brace Jovanovich, Pub, 1984.
[5] Feng H , Fu Q , Zhang L . How to Launch a New Durable Good: A Signaling Rationale for Hunger Marketing[J]. International Journal of Industrial Organization, 2020, 70.
[6] After Apple M1, Intel can no longer "squeeze toothpaste" at will[EB/OL].
https://ouranzoola.istocks.club/after-apple-m1-intel-can-no-longer-squeeze-toothpaste-at-will/2020-11-22/

**Author introduction**: **Shudong YANG**, PhD candidate at Dalian University of Technology, Research field: Educational Data Mining, Email: ysd@mail.dlut.edu.cn, ORCID: 0000-0002-5044-4078 ratchet effect[7]. **In the field of education**, middle school students hide their strengths through precise low-score control or low-ranking tactics. Scholars adjust the pace of achievement and outputs according to the cycles of professional titles and performance. it is not conducive to mid- and long-term planning when the term of the university president is short, which in turn affects university performance[8]. Under the mentor-ship system that lacks an effective incentive mechanism, the master may be worried about they would lose their job when their student master the knowledge, leading to the phenomenon of "not-do-their-best" commonly[9].

Here will give a brief definition of the above social phenomenon for the convenience of description. **Lobification** refers to a collection of performance control strategies, methods, and means that a rational people deliberately hides behaviors, or deliberately lowers performance, or deliberately delays returns for a certain long-term purpose, but without risks. In the academic field, lobification is reflected in the operation of the academic capital surplus, and it is called "**academic lobification**".

Academic lobification is an interfering factor in learning analysis, academic risk prediction, and academic achievement prediction, and needs to be compensated for bias. Academic lobification have the following application value: in terms of technology, it can improve the performance of predictive models; in terms of student management, it can mine abnormal academic behaviors to achieve "preventive treatment"; in terms of business management, it can optimize performance incentives and mechanism design .

# 2. Interdisciplinary research

## 2.1 Social psychology: explanation of the motivation for lobification

Self-worth theory believes that: "People are born with the need to maintain self-esteem and sense of self-worth"[10]. Self-disclosure and self-concealment are social control methods used to influence the development of interpersonal relationships[11]. The self-discrepancy theory believes that the greater the gap between the true self and the self-expression, the stronger the sense of security[12]. Therefore, the gap between public self-consciousness and privatte self-consciousness can be used to explain the motivation of individual behavior[13]. The combined effect of self-efficacy and the expected behavioral consequences can also be used to explain the behavior and emotional state of students[14][15].

## 2.2 Economics: basic assumptions for student behavior

New institutional economics believes that people have opportunistic tendencies and will use purposeful and strategic means to maximize their own interests, including concealing part of information, deliberately inducing and even deceiving[16]. New institutional economics also believes that people have bounded rationality. Due to the uncertain teaching environment, incomplete information, and the limited cognitive ability of students, the subjective models of students' responses to the environment are inconsistent, leading to differences in behavioral choices (Diversity, complexity)[17]. A student who can well control the "delayed gratification" can easily make a rational choice in the intertemporal choice[18], thereby increasing the possibility of long-term success.

[7] Zhang Weiying. Game Theory and Information Economics (New 1st Edition)[M]. Gezhi Press, 2012: 271-274.
[8] Liu Chang, Chen Shouming. President tenure and university performance: An empirical research based on the panel data from 1999 to 2018[J]. Science Research Management, 2019, 40(5):7.
[9] Wang Hao. The optimization method of the apprenticeship system[J]. Enterprise Management,2012(8):83-84. DOI:10.3969/j.issn.1003-2320.2012.08.027.
[10] Covington, Martin V . The Will to Learn: A Guide for Motivating Young People[M]. Cambridge University Press, 1997.
[11] Taylor S.E., Peplau L.A., Sears D.O.. Social Psychology: 10th Edition[M]. Peking University Press, 2006: 286-292.
[12] Taylor S.E., Peplau L.A., Sears D.O.. Social Psychology: 10th Edition[M]. Peking University Press, 2006: 114-115.
[13] Taylor S.E., Peplau L.A., Sears D.O.. Social Psychology: 10th Edition[M]. Peking University Press, 2006: 117-118.
[14] Bandura A.. Self-efficacy: Implementation of Control[M]. East China Normal University Press, 2003.: 27-30
[15] Stephen P. Robbins, Timothy A. Judge. Organizational Behavior[M]. Beijing: China Renmin University Press, 2008: 170-172
[16] Duan Wenbin Chen Guofu. Institutional Economics: Institutionalism and Economic Analysis[M]. Nankai University Press, 2003.

## 2.3 Game theory: the internal mechanism of behavioral choice

A study has shown that the signal transmission model can be used to explain the impact of different assessment methods on the differences in students' learning attitudes and engagement levels[19]. Game-based property rights theory believes that the choice of behavior depends on the allocation structure of residual control rights and residual claim rights[20]. Under the "Bubka Behavior Model", the perpetrator has no residual claim rights, but has complete residual control rights. As a result, he "not-do-their-best" again and again; however, under the "Jordan Behavior Model", the perpetrator has neither residual claim rights nor There is no remaining control, leading to the result of "do-their-best"[21].

## 2.4 Data science: behavioral computing

Before the popularization of smart phones, the MIT Media Lab had already started to predict behavior based on social measuring instruments by mining honest social signals[22]. Later, in the era of mobile Internet, reality mining was carried out through sensors such as wearable devices and smartphones[23][24][25][26]: relationship mining, and personal behavior pattern analysis, etc. The common methods include: behavior modeling, which uses related methodology and tools to capture and express behavior characteristics; behavior analysis, which is behavior inference based on context; behavior mining, which connects behavior patterns with personal attributes, in order to mine causality and anomaly detection[27].

## 2.5 Behavioral science: behavioral design

there are many products based on the Fogg Behavior Model[28][29] for behavior design among internet education applications[30]. A common trick is educational gamification[31]. On the offline end, a public school called "Quest to Learn" in New York, USA, uses gamification to encourage students to devote themselves to learning as if they were playing games[32].

[17] Lu Xianxiang, Zhu Qiaoling. New Institutional Economics (Second Edition)[M]. Peking University Press, 2012: 213-214, 183-185 (Kindle edition)
[18] Paul A. Samuelson. A Note on Measurement of Utility[J]. Review of Economic Studies, 1937, 4(2): 155-161.
[19] Li Jihong, Zhao Tao. The Analysis of Students' Screening Model in Optional Courses of University[J]. Chinese agricultural mechanization, 2010(6):109-112. DOI:10.3969/j.issn.1006-7205.2010.06.028.
[20] Bazel Y. Economic Analysis of Property Rights[M]. Shanghai People's Publishing House, 1997. 109-112, 117
[21] Zhang Fajun. "Remaining one hand" and "work like a dog" from time to time[J]. Enterprise Management,2009(8):23-25. DOI:10.3969/j.issn.1003-2320.2009.08.008.
[22] Pentland A S. Honest Signals: How they Shape Our World[M]. The MIT Press, 2008: 6-14 (Kindle edition).
[23] Eagle N, Pentland A S. Reality Mining: Sensing Complex Social Systems[J]. Personal and Ubiquitous Computing, 2006, 10(4):255-268.
[24] Eagle N, Pentland A S, Lazer D. Mobile Phone Data for Inferring Social Network Structure[M]. Springer US, 2008.
[25] Eagle N, Pentland A S, Lazer D. Inferring friendship network structure by using mobile phone data[J]. Proceedings of the National Academy of Sciences, 2009, 106(36):15274-15278.
[26] Pentland A S. Social Physics: How Good Ideas Spread-The Lessons from a New Science[M]. Penguin Press, 2014
[27] Cao L., Philip S. Y.. Behavior Computing: Modeling, Analysis, Mining and Decision[M]. London: Springer, 2012
[28] B. J. Fogg. Tiny Habits: The Small Changes That Change Everything[M]. Houghton Mifflin Harcourt, 2019: 1-12
[29] Fogg B J . A behavior model for persuasive design[C]// International Conference on Persuasive Technology, 2009
[30] Werbach K , Dan H . For the Win: How Game Thinking can Revolutionize your Business[M]. Wharton digital press, 2012.
[31] McGonigal J. Reality Is Broken: Why Games Make Us Better and How They Can Change the World[M]. Penguin Books, 2011: 257 (Kindle edition)
[32] The MacArthur Foundation. Quest to Learn[EB/OL]. https://www.q2l.org/

## 2.6 Pedagogy: Application areas of lobification

The discovery of academic lobification has a bias compensation effect on student evaluation: it can restore the true self of the "problem student" to solve the double misunderstanding between the student and the teacher[33]. In addition, it also has a bias compensation effect on teaching, curriculum, and teacher evaluation. The discovery of academic lobification also has a bias compensation effect on the early warning of academic risk: gifted underachievers usually do not meet the category of academic risk[34], and true academic input is the key to identifying gifted underachievers in addition to IQ[35].

# 3. Research objects and research questions

## 3.1 Related concepts

In order to explain the phenomenon of super-utilitarian capital accumulation and exchange, Bourdieu expanded capital into economic capital, cultural capital and social capital, and explained the internal mechanism of social game rules through the distribution structure of different types of capital[36]. Academic capital is inherited from cultural capital. It is a subdivided concept, which represents the ability to control academic performance. It is embodied in the form of grades and academic achievements. Academic capital can be accumulated through genetics and learning. It has the characteristics of embodiment and can be transformed into cultural abilities, cultural products and cultural qualifications.

Below will give a complete definition of academic lobification: **Academic lobification** refers to a collection of academic performance control strategies, methods, and means that a rational student deliberately hides academic behaviors, or deliberately lowers academic performance, or deliberately delays academic returns for a certain long-term purpose, but without academic risks. Academic lobification is reflected in the operation of the academic capital surplus, and is manifested in the control of academic performance leeway or margin. Generally speaking, capital is accumulative and liquid[37], however because academic capital depends on student, the surplus of academic capital does not have interpersonal mobility. The life cycle of academic lobification includes three stages: initialization period, lobification period, and exposure period. For college students, the initial period mainly refers to the period from the college entrance examination to the time before the formation of the class group. The related concepts are illustrated as follows:

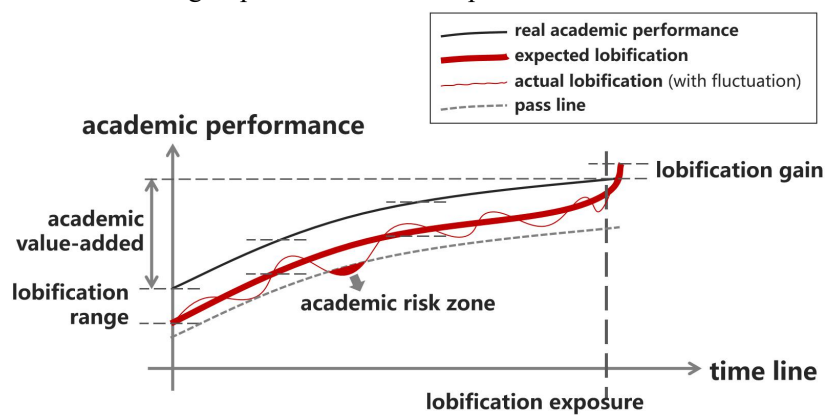

Figure 1 related concepts about academic lobification

[33] Wang Qiqiao. From Misunderstanding to Understanding: Explanation of the Meaning of "Problem Students"[J]. Teaching & Adimistration (Theory Edition), 2020(3): 62-65.
[34] Jean, Sunde, Peterson, et al. Gifted Achievers and Underachievers: A Comparison of Patterns Found in School Files[J]. Journal of Counseling & Development, 1996, 74(4):399-407.
[35] Steenbergen-Hu S., Olszewski-Kubilius P., Calvert E. The effectiveness of current interventions to reverse the underachievement of gifted students: Findings of a meta-analysis and systematic review[J]. Gifted child quarterly, 2020, 64: 132-165.
[36] Bourdieu P. Cultural Capital and Social Alchemy[M]. Shanghai People's Publishing House, 1997.
[37] Marx. Capital (Volume Two) [M]. Shanghai Sanlian Bookstore, 2009: 355-356.

Academic Lobification: Low Performance Control Strategy for Long-planed Academic Purpose

Other related concepts include: academic lobification release, that is, students stop their lobification and begin to release their academic capital surplus. The turning point from the academic lobification period to the exposure period is called "academic lobification exposure", that is, the lobification behavior is recognized or seen through. There are two situations of exposure: Situation 1, occurs after the academic lobification release, there may be a certain delay from academic lobification release to the recognition. Such exposure is the deliberate behavior of academic lobification; Situation 2, occurs before the academic lobification release, that is, the behavior of academic lobification is detected in advance, and such exposure is the unintentional behavior of academic lobification. The gain of academic lobification is the extra benefit brought by the extraordinary performance due to the sense of security generated during the lobification period.

## 3.2 Research object

The objects of academic lobification research include but are not limited to: middle school students, college students, graduate students, post-doctoral fellows, scholars (young teachers, tenured faculty members), etc.

## 3.3 Research question

Question classification includes two dimensions: micro or macro, dynamics or statics. From the micro perspective: How does a specific individual gain academic lobification? From the macro perspective: How do specific groups interact the game of academic lobification? From the perspective of statics: what are the elements, structure, and mechanism of academic lobification? From the dynamics perspective: what is the trend of change and development of academic lobification? For detail as follows:

- **The mechanism of academic lobification**(micro + statics), that is, the study of behavioral motivations and external incentives , which has individual contingency and unpredictability.
- **Individual development of academic lobification**(micro + dynamics), that is, the study of the trend of individual academic lobification with the growth of grades.
- **Academic interaction networks**(macro + statics), that is, studying non-cooperative game and equilibrium conditions, analysis and application of academic interaction network structure, which has group inevitability and predictability.
- **The group evolution of academic lobification**(macro + dynamics), that is, studying the evolution prediction and application of academic interaction networks.

# 4. Research scope

## 4.1 Individual behavior

**1) The hiding of academic behavior: pretending not to study hard, bluffing**
American psychologist Martin Covington's Self-worth Theory believes that "people are born with the need to maintain self-esteem and sense of self-worth." Some students may reserve margin for themselves or stakeholders to explain their academic failures in advance. Students will pretend not to study hard and set up self-barriers to protect themselves. In China, this phenomenon is common among middle school students, undergraduates and graduate students who are competing for scholarships or further studies.

**2) Academic low-performance control: scores, rankings, and echelons**
For middle school students, academic low-performance control may be motivated by: lowering scores in the usual exams

Academic Lobification: Low Performance Control Strategy for Long-planed Academic Purposecan make them stable and emotional in the final exam, so that they can perform supernormally; in order to show a state of "continuous progress", so as to be able to obtain material or spiritual rewards from teachers or parents; to maliciously paralyze competitors; to avoid being harassed by slacker students during exams; to avoid appearing under the spotlight of the crowd, etc.

For college students, the "input-output theory" believes that the ratio of input to output has a diminishing marginal effect. For college students/graduates who do not pursue short-term benefits such as scholarships, they only need to pursue a GPA2.0 or pass, and the remaining limited energy is focused on more important things.

Such behavior has an interference effect on teaching evaluation and student evaluation. For example, gifted students do not necessarily have good grades; high-performing students do not necessarily have the potential for future development. Therefore, research scopes include: bias compensation of teaching and student evaluation, gifted student mining, etc.

**3) Academic delayed gratification: Slowing down for academic achievement, paper inventory control**
The hypothesis of the "rational man" in economics holds that the economic actions taken by every person engaged in economic activities are trying to obtain their own maximum economic benefits at their own minimum economic costs. According to external variables such as academic stage, graduation requirements, performance cycle, students control the delayed pace of academic achievement output in a planned way to maximize personal benefits. This phenomenon is common in graduate students, post-doctoral fellows, and young university teachers. Therefore, research scopes include but not limited to: Slowing down for academic achievement, paper inventory control, Paper publication frequency and phase control, etc.

## 4.2 Group behavior

The research scope of group behavior includes, but is not limited to, exposure and perception of academic lobification, non-cooperative game and equilibrium, academic interaction network, academic lobification and group subculture. Among them, the non-cooperative game and equilibrium are further subdivided into: incomplete information and reputation mechanism, asymmetric information and signal transmission mechanism, etc.

## 5. Research methods

The validity and real-time performance of academic lobification inference are restricted by three forces: data collection ability, algorithm performance, and computing power. With the development of Internet of Things technology, data science, computer hardware, student mobile phone dependence, and the popularization of college informatization, The above three forces are no longer the bottleneck. The calculation framework is shown in the figure below:

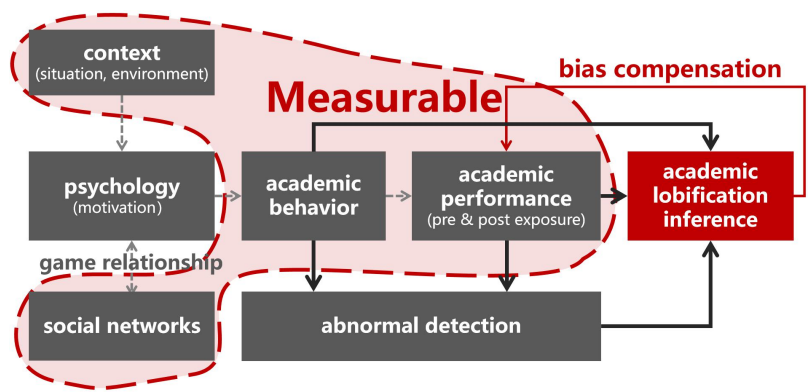

Figure 2 The calculation framework for academic lobification inference

## 5.1 Basic work

The academic lobification inference depends on the collection of mobile perception data. Currently, the mature collection frameworks mainly include the Funf framework, the Device Analyzer framework, and the UbiqLog framework. The former is an extensible sensing and data processing framework based on Android, developed by the MIT Media Lab. Funf provides a set of reusable functions that support the collection, upload and configuration of a wide range of data types[38][39].

The academic lobification inference requires the support of data integration technology and data fusion technology to integrate mobile perception data and school management system data. After that, data cleaning and data preprocessing are performed by means of data noise reduction, missing value completion, structuring, time slicing, and spatial gridding.

## 5.2 Academic behavior computing

Academic behavior needs to be computed according to the context . The main work of behavioral context includes event definition, that is, the design of variables, and state definition, that is, the combination of various events: actions, time slices, spatial grids, environmental conditions, etc. Then conduct behavior inference based on context awareness, including three levels respectively: establish the mapping rules of state and behavior, divide normal behavior and abnormal behavior; calculate the distribution of individual states(In other words, where did each student's time go?); Calculate the distribution of group status.

## 5.3 Academic performance computing

Academic performance needs to be computed based on the student's user portrait model and time series model. Among them, the user portrait training includes data initialization (entrance exams, etc.), personality trait evaluation (MBTI, etc.), and real-time dynamic adjustment of user portraits based on academic performance. The time series model mainly trains personal academic performance development.

## 5.4 Academic lobification inference

Regardless of whether the lobification is exposed or not, it can be judged by the degree of difference between academic behavior and academic performance. The main work of this part is behavior-performance mapping training, unsupervised anomaly detection, including abnormal behavior discovery and abnormal performance discovery.

The above calculations are inseparable from machine learning, but general machine learning is not interpretable. However, it can be explained by feature attribution methods[40][41] (such as SHAP) or counterfactual interpretation methods[42] (such as DiCE).

---

[38] Aharony N , Pan W , Ip C , et al. Social fMRI: Investigating and shaping social mechanisms in the real world[J]. Pervasive & Mobile Computing, 2011, 7(6):643-659.
[39] Alan. G. funf-core-android[EB/OL]. Github, https://github.com/funf-org/funf-core-android
[40] S Masís. Interpretable Machine Learning with Python: Learn to build interpretable high-performance models with hands-on real-world examples[M]. Packt Publishing, 2021.
[41] C Molnar. Interpretable Machine Learning: A Guide for Making Black Box Models Explainable[M]. lulu.com, 2020.
[42] Pearl J, Mackenzie D. The book of why: the new science of cause and effect[M]. Beijing: CITIC Press, 2019

# 6. Discussion and outlook: how will academic lobification develop?

In order to maximize the potential benefits of academic lobification for education management, we must understand the individual academic lobification behavior and motivations, as well as the possible impact on the group. To this end, we need a new interdisciplinary research field: **Lobificationlogy**. In order to make this field develop smoothly, the factors that must be considered in the following aspects:

- Data is the foundation. Explaining any behavior cannot be completely separated from environmental data. Therefore, it is necessary to integrate broader environmental data.
- Multidisciplinary efforts are required, and the research process is accompanied by challenges brought about by interdisciplinary cooperation. Meeting these challenges is crucial. Universities, governments, and funding agencies can play an important role in promoting interdisciplinary research.
- Research on student-related topics may bring legal and ethical issues to researchers. Therefore, it is necessary to circumvent legal, ethical, and ethical issues through big data governance and other means.